\documentclass[twocolumn,showpacs,aps,groupedaddress,prd,eqsecnum,epsf]{revtex4}
\usepackage{latexsym,amsmath,graphicx}

\usepackage{epsfig}
\usepackage{dcolumn}
\usepackage{bm}
\usepackage{latexsym}
\usepackage{amsfonts}
\usepackage{amssymb}

\newcommand{\beq}{\begin{equation}}
\newcommand{\eeq}{\end{equation}}
\newcommand{\beqa}{\begin{eqnarray}}
\newcommand{\eeqa}{\end{eqnarray}}
\newcommand{\bea}{\begin{array}}
\newcommand{\ena}{\end{array}}

\begin{document}
\draft

\title{Continuous area spectrum in regular black hole}

\author{Hidefumi Nomura$^{1}$}
\email{nomura@gravity.phys.waseda.ac.jp}
\author{Takashi Tamaki$^{2}$}
\email{tamaki@tap.scphys.kyoto-u.ac.jp}
\address{$^{1}$Department of Physics, Waseda University, 3-4-1 Okubo,
Shinjuku, Tokyo 169-8555, Japan~ \\
$^{2}$Department of Physics, Kyoto University, 
606-8501, Japan~}

\date{\today}

\begin{abstract}
We investigate highly damped quasinormal modes of
regular black hole coupled to nonlinear electrodynamics. 
Using the WKB approximation combined with complex-integration 
technique, we show that the real part of the frequency disappears 
in the highly damped limit. If we use the Bohr's correspondence principle, 
the area spectrum of this black hole 
is continuous. We discuss its implication in the 
loop quantum gravity. 
\end{abstract}

 \maketitle

\section{Introduction}

Quasinormal modes (QNMs) of black holes are solutions of the classical
perturbation equations in the gravitational background with 
the specific boundary conditions for purely outgoing waves
at infinity and purely ingoing at the event horizon.
QNMs are intrinsic quantities that characterize black holes. 
These are expected to dominate the emitted radiation in many dynamical processes 
involving a black hole at late times. Since QNMs are also expected to reveal the 
information about parameters of black holes, it is important in detecting them  
from the astrophysical viewpoint. For a review, see, e.g., \cite{Kokkotas}. 

During the last few years,
they have also attracted much attention in the context of quantum gravity. 
This is related to the area quantization of black holes discussed by
Bekenstein \cite{Bekenstein}. First, we identify the real part of the 
highly damped QNMs as a minimum change of the black hole 
mass based on the Bohr's correspondence principle \cite{Hod}. 
For Schwarzschild black hole, we have \cite{Motl}
\begin{equation}
{\rm Re}(\omega)=T_H {\rm ln}3 \qquad {\rm as}  \quad |{\rm Im} (\omega)|
\rightarrow \infty\ .
\end{equation}
Applying the first law of black hole thermodynamics, we obtain
\begin{equation}
dA=4 dM/T_H=4 {\rm ln}3 \label{SchQNM}
\end{equation}
where $dM=dE={\rm Re}(\omega)$. 
The reason why it has been paid attention is the relation to the 
loop quantum gravity where the area spectrum is 
given by \cite{Rovelli} 
\beqa
A=8\pi \gamma \displaystyle  \sum_{i} \sqrt{j_i(j_i+1)}
\ , \label{AreaLQG}
\eeqa
where $\gamma$ is the Immirzi parameter related to an ambiguity 
in the choice of canonically conjugate variables \cite{Immirzi}. 
The sum is added up all intersections between a surface and a spin 
network carrying a label $j=0$, $1/2$, $1$, $3/2$, $\ldots$ reflecting 
the SU(2) nature of the gauge group. 
The statistical origin of the black hole entropy $S$ is also derived in 
\cite{Ashtekar}. The idea is to 
identify the minimum element in (\ref{AreaLQG}) $A_{\rm min}$ and 
(\ref{SchQNM}), i.e.,  
\beqa
dA=4\ln 3=8\pi \gamma\sqrt{j_{\rm min}(j_{\rm min}+1)}\ , 
\label{identify}
\eeqa
and use the relation $S=A/4$ \cite{Dreyer}. Then, 
$j_{\rm min}$ is determined as $1$, which is consistent with the requirement 
that $j$ is half-integer. Since this consistency seems meaningful, 
various arguments has been done \cite{Alekseev,Corichi2,Khriplovich,
Kaul,Corichi3,Ling}. 

The main reasons opposing this idea are summarized as follows. 
(i) Other black holes, such as 
Reissner-Nordstr\"om black hole \cite{Motl,Andersson,Berti}, Schwarzschild 
de-sitter (dS) 
black hole \cite{Cardoso,Shijun},  Kerr black hole \cite{Shijun2},
and also in d-dimensional  
Schwarzschild and Reissner-Nordstr\"om with a cosmological constant \cite{Schiappa}
do not have above consistency. (ii) Original calculation of the black hole 
entropy has a mistake \cite{Domagala,Meissner}. The corrected entropy 
suggests that $j_{\rm min}$ detemined above way is not half-integer. 

On the other hand, there are also reasons supporting this idea. 
(i) Schwarzschild black hole in other dimensions has 
the relation (\ref{SchQNM}) \cite{Motl,Kuns,Birm,Lemos}. Surprisingly, single-horizon 
black holes, such as dilatonic black hole \cite{GM-GHS} shares the 
relation (\ref{SchQNM}) \cite{Tamaki} as it has been suggested in \cite{Visser,Pad}. 
This has been confirmed in other way in \cite{Kettner,Das}. This universality suggests 
meaningful. (ii) The black hole entropy has been reexamined based on the idea that 
spherical symmetry should be reflected in the number counting of microstates 
for spherically symmetric black holes \cite{Dreyer2}. In this case, original 
consistency that $j_{\rm min}=1$ has been recovered.   

As shown above, this is still controversial. Therefore, we need further 
discussion and study from both QNMs and the loop quantum gravity. 
It is also interesting that QNMs of AdS black holes have 
a direct interpretation in terms of the dual conformal 
field theory (CFT) \cite{Horo,Bir1} according to the AdS/CFT correspondence 
\cite{Mal,Witten}. There is also a possibility that QNMs of AdS black holes 
play an important role in determining the microstates of black holes 
\cite{Chen}. Its application to the general case is still hypothetical 
\cite{Bir2}. However, it is stimulating that two candidates of the quantum 
gravity suggest the important role in QNMs independently. 

Therefore, it is natural in examining QNMs  of black holes with 
quantum gravity motivated model, for example, Gauss-Bonnet black hole \cite{Konoplya}. 
In such a model, it is expected to 
be singularity-free. In this paper, 
we focus on the highly damped QNMs of
``regular" (no singularity inside the horizon)
 black hole coupled to the nonlinear electrodynamics 
satisfying the weak energy condition \cite{Eloy,Breton}. 
We also discuss its interpretation from the loop quantum gravity.

\section{Highly damped QNMs of Regular Black Hole}

Here, we investigate the asymptotic QNMs of
regular black hole using the WKB analysis combined with 
complex-integration technique following \cite{Andersson}. 
We use the line element of the regular black hole obtained in 
Einstein gravity coupled with nonlinear 
electrodynamics proposed in \cite{Eloy} which can be expressed as
\begin{eqnarray}
\hspace{-10mm}&&ds^2=-\left(1-\frac{2Mr^2}{(r^2+q^2)^{3/2}}+\frac{q^2r^2}{(r^2+q^2)^2}
\right)dt^2
\nonumber \\
\hspace{-10mm}&&+\left(1-\frac{2Mr^2}{(r^2+q^2)^{3/2}}+\frac{q^2r^2}{(r^2+q^2)^2}
\right)^{-1}dr^2+r^2d\Omega^2, \label{eq:metricrbh}
\end{eqnarray}
where the associated electric field $E$ is given by 
\begin{equation}
E=q\,r^4\left( \frac{r^2-5\,q^2}{(r^2+q^2)^4}+\frac{15}2\,\frac M{%
(r^2+q^2)^{7/2}}\right) .  \label{eq:E}
\end{equation}
Note that this solution asymptotically behaves as the
Reissner--Nordstr\"om solution, 
\begin{eqnarray}
-g_{tt}&=&1-2M/r+q^2/r^2+O(1/r^3), \nonumber \\
      E&=&q/r^2+O(1/r^3).
\end{eqnarray}
Thus, the parameters $M$ and $q$ are related 
correspondingly with the mass
and the electric charge. 
If $|q|<2s_cM$ ($s_c \simeq 0.317$ see \cite{Eloy}), this expresses 
a regular charged black hole which has inner horizon $r_{-}$
and event horizon $r_{+}$. 
We concentrate on this solution from now on.
We define
\begin{eqnarray}
g(r) &\equiv & 1-\frac{2m(r)}{r},  
\end{eqnarray}
where
\begin{equation}
 m(r)= \frac{Mr^3}{(r^2+q^2)^{3/2}}-\frac{q^2r^3}{2(r^2+q^2)^2}.
\end{equation}
Notice the relation to the Hawking temperature $T_{\rm H}$ is \cite{Visser}
\begin{eqnarray}
g'(r_{+})=4\pi T_{\rm H}, \label{H-T}
\end{eqnarray}
where $':=d/dr$. 
The perturbation (Regge-Wheeler) equation for (\ref{eq:metricrbh}),
 with the time dependence $\exp(-i\omega t)$,
is 
\begin{eqnarray}
\frac{d^2 \psi}{dr_{\ast}^{2}}+[\omega^2 -V(r)]\psi=0\ , \label{Schro}
\end{eqnarray}
where $r_*$ denotes the tortoise coordinate given by
\begin{eqnarray}
\frac{dr_{\ast}}{dr}=\frac{1}{g(r)}\ .
\end{eqnarray}
and the Regge-Wheeler potential is given by \cite{Visser}
\begin{eqnarray}
V(r)&=&g \left( \frac{l(l+1)}{r^2}+(1-k^2)\frac{2m}{r^3} \right. \nonumber \\
    && \left. +(1-k)(\frac{g'}{r}-\frac{2m}{r^3}) \right)
\end{eqnarray}
$k=0$, $1$, and $2$ for scalar, electromagnetic,
 and odd parity gravitational
perturbations, respectively.
We impose the boundary consitions, which are
purely outgoing plane
waves at spatial infinity and purely ingoing plane waves
at the horizons, on $\psi$ later. 
Introducing $\Psi=g^{1/2}(r)\psi$, we can rewrite 
(\ref{Schro}) as
\begin{eqnarray}
\frac{d^2\Psi}{dr^2}+R(r)\Psi=0, \label{eq:p-eq1}
\end{eqnarray}
where 
\begin{eqnarray}
R(r)&=& g^{-2}\left( \omega^2 - V(r)+\frac{{g'}^2}{4}-\frac{gg''}{2} \right).
\end{eqnarray}

>From now, we consider the WKB analysis combined with 
complex-integration technique, which is a good approximation
in the limit ${\rm Im}(\omega)\rightarrow -\infty$.
We seek for the WKB condition which corresponds to 
the monodromy condition of Motl and Neitzke \cite{Motl}.
The two WKB solutions to an equation of 
form (\ref{eq:p-eq1}) can be written as 
\begin{eqnarray}
\Psi_{1,2}^{(s)} (r)=Q(r)^{-1/2}\exp
\left( \pm i\int_{s}^{r} Q(r')dr'\right), \label{eq:WKB}
\end{eqnarray}
with $Q^2\equiv R+ ({\rm extra\ term})$.
The zeros and poles of the function $Q^2$ play a central role
in our complex analysis \cite{Andersson}. Notice that the zeros 
approach $r^2 \simeq -q^2$ in the limit ${\rm Im}(\omega)\rightarrow -\infty$.   
Here, We choose the (extra term) for $\Psi$ to behave properly near 
$r^2 \simeq -q^2$.

Expanding near $r^2 \simeq -q^2$, we obtain 
\begin{eqnarray}
R(r) &\simeq& \frac{(r^2+q^2)^4}{q^8}(\omega^2 -
 8\frac{q^{4}r^6}{(r^2+q^2)^6}).
\end{eqnarray}
It is independent of $k$.
Therefore, the perturbation equation near $r^2\simeq -q^2$ becomes
\begin{eqnarray}
\frac{d^2\Psi}{dr^2}+\frac{8q^2}{(r^2+q^2)^2}\Psi=0.
\label{eq:p-eq}
\end{eqnarray}

Thus, the asymptotic solution can be written as 
\begin{eqnarray}
\Psi &\simeq& (r \mp iq)^{1/2 \pm 3/2} 
\qquad (r\simeq \pm iq). \label{eq:asymp-sol}
\end{eqnarray}
Then we should choose
\begin{eqnarray}
Q^2&=&R+\frac{q^2}{(r^2+q^2)^2}  \nonumber  \\
&\simeq &\frac{(r^2+q^2)^4}{q^4r^4}
(\omega^2-9\frac{q^4r^6}{(r^2+q^2)^6})
\end{eqnarray}
for the WKB solution (\ref{eq:WKB})
to coincide with (\ref{eq:asymp-sol}) near $r\simeq \pm iq$. 
This is analogous to the``Langer modification" 
$l(l+1) \rightarrow (l+1/2)^2$
that is used in the WKB analysis of radial quantum problems \cite{Langer}.

\begin{figure*}[tp]
\psfig{file=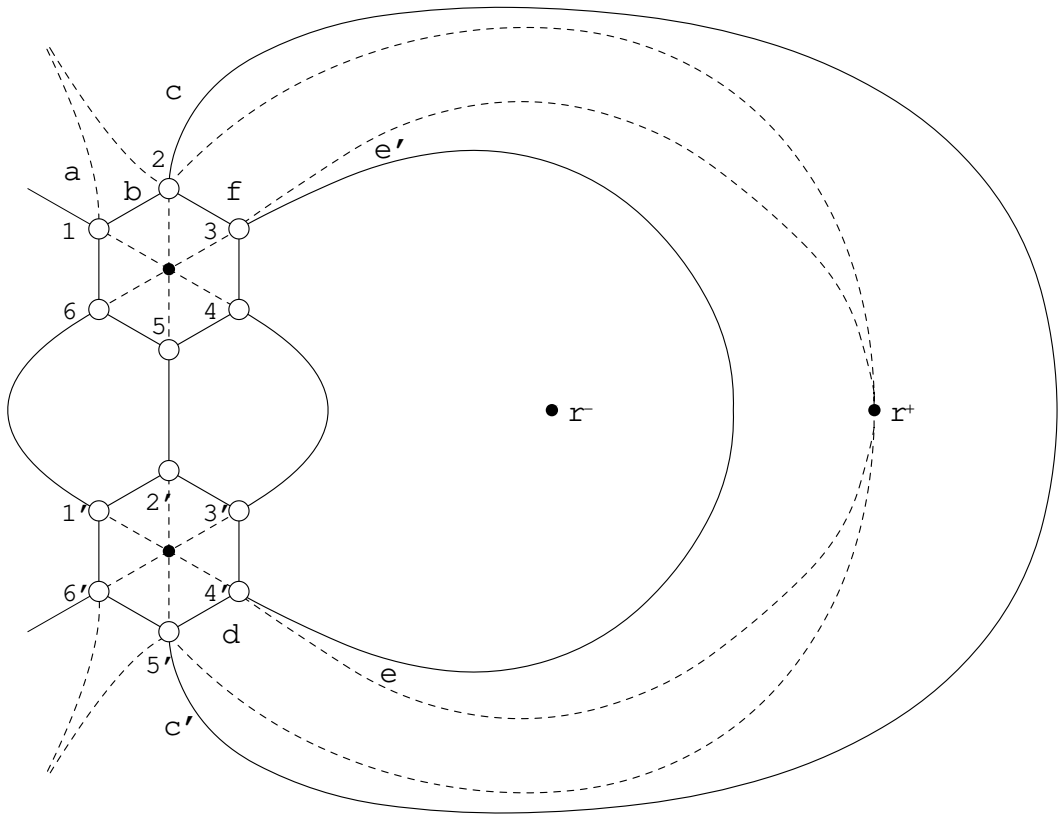,width=5.5in}
\caption{A schematic illustration of the Stokes (dashed) and
anti-Stokes (solid) lines for the regular black hole problem in the 
complex $r$-plane in the limit Im$(\omega )\to -\infty$.
The open and filled circles represent zeros and poles of $Q^{2}(r)$
respectively. 
 \label{Stokes} }
\end{figure*}

Then, we can find that the function $Q^2$ has four second order poles 
($r=r_{-}$, $r_{+}$,  $\pm iq$) and twelve zeros 
(around $r= \pm iq$). 
We depict these poles and zeros in Fig.~\ref{Stokes}.
We explain the technique that is crucial 
for our analysis. From each simple zero of $Q^2$ emanates three so-called 
``Stokes lines''. 
Along each of these contours, $Q(r) dr$ is purely imaginary, 
which means that one of the two solutions grows exponentially
while the second solution decays, as we move this line. 
In other words, one of the solutions is exponentially dominant
on the Stokes line, while the other solution is sub-dominant.
Analogously, one can define ``anti-Stokes lines'' associated with each 
simple zero
of $Q^2$. On anti-Stokes lines,  $Q(r) dr$ is purely real, 
which means that the two solutions are purely oscillatory. 
As we cross an anti-Stokes line, the dominancy of the two 
functions $\Psi_{1,2}$ changes.

Stokes lines are vital for WKB analysis, because the solution changes 
character in the vicinity of these contours. That is, if the solution 
is appropriately represented 
by a certain linear combination of $\Psi_1$ and 
$\Psi_2$ in some region of the complex 
$r$-plane, the linear combination will change as the solution is 
extended across a Stokes line. The induced change is not 
complicated: The coefficient of the dominant solution remains 
unchanged, while
the coefficient of the other solution picks up a contribution proportional
to the coefficient of the dominant solution.
This is known as the ``Stokes phenomenon'' \cite{Stokes}.
The constant of proportionality is known as a ``Stokes constant''. 
This change is necessary for the particular representation (\ref{eq:WKB}) 
to preserve the monodromy of the global solution. Terms that are exponentially 
small in one sector of the complex plane may be overlooked. However, in 
other sectors they can grow exponentially and  
dominate the solution. By incorporating the
Stokes phenomenon, we have a formally exact procedure which leads to 
a  proper account of all exponentially small terms.

In the particular case of an isolated simple zero of $Q^2$ the problem is 
straightforward. We choose the phase of the square-root of $Q^2$ such that 
\begin{equation}
Q = R^{1/2} \sim \omega \quad \mbox{ as } r\to \infty\ .
\end{equation}
This means the boundary conditions which are
 the outgoing-wave solution at infinity is proportional to 
$\Psi_1$ while the ingoing-wave solution at the horizon is 
proportional to $\Psi_2$. 
Suppose that the solution in the initial region of
the complex plane is given by
\begin{equation}
\Psi = c\Psi_1^{(s)}\ .
\end{equation}
Then, after crossing a Stokes line
emanating from $s$ (and on which $\Psi_1$
 is dominant) the solution becomes
\begin{equation}
\Psi =   c \Psi_1^{(s)} \pm i c \Psi_2^{(s)}\ .
\end{equation}
The sign  depends on whether
one crosses the Stokes line in the positive (anti-clockwise) or 
negative (clockwise) direction.
It is crucial to note that this simple result, i.e. that the Stokes constant
is $\pm i$, only holds when the Stokes line 
emanates from the zero that is  used as lower limit for the phase-integral. 
That is, when we want to 
use the above result to construct an approximate solution valid in various 
regions of the complex plane, we often change the reference point 
for the phase-integral. In this case, it is necessary to evaluate integrals 
of the type
\begin{equation}
\gamma_{ij} = \int_{s_i}^{s_j} Q(r) dr\ ,
\end{equation}  
where ${s_i}$ and ${s_j}$ are two simple zeros of $Q^2$. 

Let us evaluate the above integrals.
Near $r\simeq  \pm iq$, we evaluate the phase-integrals
\begin{eqnarray}
\hspace{-10mm}&&I \equiv  \int Qdr\ , \nonumber \\
\hspace{-10mm}&&\simeq \pm \int \frac{(r^2+q^2)^2}{q^4}\left(\omega^2+9\frac{q^{10}}
{(r^2+q^2)^6}\right)^{1/2}dr\ , \nonumber \\
\hspace{-10mm}&&\simeq \mp
          \int \frac{4(r\mp iq)^2}{q^2}
      \left(\omega^2-\frac{9}{64}\frac{q^4}{(r\mp iq)^6}\right)^{1/2}dr\ . 
\end{eqnarray}
If we define,  
\begin{equation}
y= \frac{8\omega(r\mp iq)^3}{3q^2},
\end{equation}
the zeros of $Q$ map to $-1$ or $1$, and we can get
\begin{eqnarray}
I &=& \mp \frac{1}{2}\int (1-\frac{1}{y^2})^{1/2}dy\ , \nonumber \\
  &=& \pm \frac{\pi}{2}\ .
\end{eqnarray}

Then, we obtain
\begin{eqnarray}
\gamma&=&-\gamma_{12}\ =\ -\gamma_{32}\ =\ \gamma_{43}\ =\ -\gamma_{54}\ ,
   \nonumber \\ 
   &=& -\gamma_{1'2'}=-\gamma_{3'2'}=\gamma_{4'3'}=-\gamma_{5'4'}\ ,
    \nonumber \\ 
  &=& \pi/2\ ,
\end{eqnarray}
where the lower indices are related to the zeros in Fig.~\ref{Stokes}.

Now we compute the QNMs utilizing ``Stokes phenomenon".
For frequencies $|\mbox{Im }\omega| \gg |\mbox{Re }\omega|$,  
the pattern of Stokes and anti-Stokes lines  
is sketched in Fig.~\ref{Stokes}. 
Assuming that Re~$\omega M >0$ the outgoing wave boundary condition 
at spatial infinity
can be analytically continued to the anti-Stokes line labeled 
$a$ in the figure.  
The method to obtain these lines is discussed in more detail, 
for example, \cite{pimrev}.
In order to obtain the WKB condition for highly damped QNMs,
we analytically continue the solution along a closed path 
encircling the pole at the event horizon.  This contour 
starts out at
$a$, proceeds  along anti-Stokes lines and account
 for the Stokes phenomenon
associated with the zeros 
 and eventually ends up at $a$. 
In other words, we choose the path
  $a \rightarrow b \rightarrow c \rightarrow c' \rightarrow
 d \rightarrow e \rightarrow e' \rightarrow f 
 \rightarrow b \rightarrow a$ in Fig.~\ref{Stokes}. 
We can arrrive at the following WKB condition for highly
damped QNMs
\begin{equation}
e^{2i \Gamma_+}=1-(1+e^{-2i\gamma})(1+e^{2i\gamma})
(1+e^{-2i\Gamma_-}) \label{eq:WKB-cond}
\end{equation}
(We can perform the calculation quite analogous to the case in the 
Reissner-Nordstr\"om black hole explained in the Appendix of \cite{Andersson}.),
where $\Gamma_+$ and $\Gamma_-$ denote the integral along a contour that
encircles, in the negative direction, the pole at $r_+$ and $r_-$ respectively.
Substituting $\gamma=\pi/2$ into (\ref{eq:WKB-cond}), we can get
\begin{equation}
e^{2i \Gamma_+}=1.
\end{equation}
Note that this expression is independent of $\Gamma_-$
unlike the Reissner-Nordstr\"om case.
This means that inner horizon does not contribute
to the QNMs.

We evaluate the integral $\Gamma_+$ using the residue theorem
\begin{eqnarray}
\Gamma_+&=&-2\pi i {\rm Res} Q(r_+)\ , \nonumber \\ 
        &=& -2\pi i \lim_{r \to r_+} 
                   \frac{r-r_+}{g(r)}\omega 
                   \sqrt{1+\frac{g'(r)^2}{4\omega^2}}\ , \nonumber \\
        &=& -2\pi i \lim_{r \to r_+} 
                   \frac{r-r_+}{g(r)}\omega\ ,  \nonumber \\
        &=& -2\pi i \lim_{r \to r_+} \frac{1}{g'(r)}  \omega\ ,   \nonumber \\ 
        &=&-\frac{\omega}{2T_H}i\ ,
\end{eqnarray}
where we use $|\omega| \rightarrow \infty$ and (\ref{H-T}).
Then WKB condition can be written as 
\begin{equation}
e^{\omega/T_H}=1\ .
\end{equation}
We can get immediately
\begin{equation}
\omega=0- i \cdot 2n\pi T_H \qquad n \rightarrow \infty.
\end{equation}

Note that this result apply to scalar, electromagnetic,
and odd parity gravitational perturbations.


\section{conclusions and discussion}
Our results show that the real part of the frequencies
is zero in the highly damped limit.
What does this mean?
Dreyer identified $dA$ of (\ref{SchQNM}) with $A_{\rm min}$ in (\ref{AreaLQG})
obtained from loop gravity.
However, the original Bohr's correspondence principle
is as follows: ``transition frequencies at large quantum numbers should equal
classical oscillation frequencies".
In other words, the distance between two 
neighboring energy levels $\Delta E$ with large quantum numbers 
(between levels with $n$ and $n+1$ ($n \gg 1$)) 
is related to the classical frequency $\omega$ in the system 
by the relation $\Delta E= \hbar \omega$.
Since the area spectrum of loop gravity is given by
(\ref{AreaLQG}),
spacing of the neighboring area spectrum $\Delta A$, in general, 
approaches zero asymptotically 
in the classical limit (when $A$ is sufficiently large). 
>From the first law of black hole thermodynamics (\ref{SchQNM}), 
it seems that the vanishing real part of $\omega$ 
supports continuous area spectrum 
applying to the original correspondence principle. 
In another context, Alexandrov and Vassilevich discuss that continuous area
apectrum follows from the Lorentz covariant loop
quantum gravity \cite{Alexandrov}.

Let us interpret the QNMs for Schwarzschild black hole or other 
single-horizon black holes in this context. It has been discussed 
that one should take into account only 
the states with the minimal spin at the horizon
counting of black hole states \cite{Dreyer2}. In this case, 
spacing of neighboring area spectrum $\Delta A$ does not approach 
zero and coincide with $A_{\rm min}$. If  this idea applies to
single-horizon black holes only, 
we can interpret the regular black hole and single-horizon black holes 
simultaneously. 
However, since it is still difficult to understand 
the results of the Reissner-Nordstr\"om and Kerr black holes,  
it is too early to conclude. Moreover, 
our analysis is only one example of QNMs of regular solutions.
The absence of $r=0$ singularity may cause the existence of zero 
real part of $\omega$ \cite{Brink}.
Therefore, we need to investigate other regular solutions.
Of course, it is also important to reconsider the number counting 
of horizon states \cite{Domagala,Meissner,Khriplovich2}. 

The subject is still debatable from a QNMs viewpoint.
QNMs boundary conditions are stated in terms of behavior of
perturbations at the horizon and infinity.
It is somewhat strange that black hole quantization
should care about infinity.
Birmingham and Carlip show that these boundary conditions for 
the BTZ black hole can be recast in terms of monodromy 
conditions at the inner and outer horizons and define 
a set of ``non-QNMs" for the higher-dimensional black holes 
involving only these monodromies \cite{Bir3}.
The correspondence principle leads to the correct quantization
of the near-horizon Virasoro gererators.
Boundary conditions for the inner horizon and outer horizon
might be the key to solve the problem.
This gives a suggestion that
single-horizon black holes have consistency, 
while dual horizon black holes, such as 
Reissner-Nordstr\"om and Kerr black holes, have no
consistency.
We need further discussion from various points of view.

\acknowledgements
We would like to thank Shijun Yoshida
and Ryo Miyazaki for useful comments
and help.
This work was partially supported by The 21st Century COE
Program (Holistic Research and Education Center for 
Physics Self-Organization Systems) at Waseda University.
This work was supported in part by Grant-in-Aid for Scientific Research Fund
of the
Ministry of Education, Science, Culture and Technology of Japan, 2003, No.\
154568
(T.T.).  This work was also supported in part by a
Grant-in-Aid for the 21st Century COE ``Center for Diversity and
Universality in
Physics".

%

\end{document}